\documentclass[12pt,reqno,a4paper]{amsart}
\usepackage[margin=1in]{geometry}
\RequirePackage{amsthm,amsmath,amsfonts}
\RequirePackage[numbers]{natbib}
\RequirePackage[colorlinks,citecolor=blue,urlcolor=blue]{hyperref}
\usepackage{booktabs,longtable}
\usepackage{graphicx}
\usepackage[table]{xcolor}
\usepackage{epstopdf}
\usepackage{physics}
\usepackage{bm}

\usepackage{setspace}
\usepackage{bm}
\usepackage[ruled,linesnumbered]{algorithm2e}
\usepackage{soul} 
\usepackage{amsaddr}
\definecolor{beaublue}{rgb}{0.74, 0.83, 0.9}
\definecolor{bananamania}{rgb}{0.98, 0.91, 0.71}

\DeclareMathOperator*{\argmin}{\arg \, \min}

\newcommand{\pkg}[1]{{\normalfont\fontseries{b}\selectfont #1}}
\let\proglang=\textsf

\begin{document}
	
\title[Clustering US States by COVID-19 Counts with NMF]{Clustering 
US States by Time Series of COVID-19 New 
		Case Counts
		with Non-negative Matrix Factorization}
	
	\author{Jianmin Chen}
	\author{Jun Yan}
	\address{Department of Statistics, University of 
	Connecticut, Storrs, CT 06269}
	\author{Panpan Zhang}
	\address{Department of Biostatistics, 
	Epidemiology and Informatincs, University of Pennsylvania, 
	Philadelphia, PA 19104}
	\email{panpan.zhang@pennmedicine.upenn.edu}
		
	\date{\today}
	
	\maketitle{}

\begin{abstract}
	The spreading pattern of COVID-19 differ a lot across
	the US states under different quarantine measures and reopening
	policies. We proposed to cluster the US states into distinct
	communities based on the daily new confirmed case counts via a
	nonnegative matrix factorization (NMF) followed by a $k$-means
	clustering procedure on the coefficients of the NMF basis. A
	cross-validation method was employed to select the rank of the 
	NMF.
	Applying the method to the entire study period from
	March 22 to July 25, we clustered the 49 continental states
	(including District of Columbia) into 7 groups, two of which
	contained a single state. To investigate the dynamics of the
	clustering results over time, the
	same method was successively applied to the time periods with
	increment
	of one week, starting from the period of March~22 to March 28. 
	The
	results suggested a change point in the clustering in the week
	starting on May 30, which might be explained by a combined 
	impact of
	both quarantine measures and reopening policies.
\end{abstract}


Secondary: 62M10 


\section{Introduction}

The COVID-19 pandemic
has brought a great deal of challenges not only to the global public
health system but also to the global economy. By July 25, 2020,
there had been a total of 15,590,252
confirmed cases across the world, while the number of death cases
had reached 630,294. The United States (US), one of the countries
that have been most severely attacked by COVID-19, reported
4,009,808 confirmed cases and 143,663 death cases, respectively, by
then (\url{https://covid19.who.int}). In the meantime, the economic
crisis caused by the pandemic was unprecedented in its scale.
The Missouri Economic Research and Information Center reported that
the total value of exports of all 50 continental US states had fallen
by 29.8\%, from \$414.95 billion in the second quarter of 2019 to
\$291.47 billion in the second quarter of 2020
(\url{https://meric.mo.gov/data/us-state-exports}).
The unemployment rate from the US Department of Labor
reached 14.7\% in April, 2020, doubling that reported
at the end of 2008 (7.2\%) when the subprime mortgage crisis broke
out. Although the unemployment rate declined to 10.2\% in July, 2020,
thanks to the stimulating measures by the government, it remained at
a high level.

Updated statistics on the pandemic and government policies are of
genuine and critical public concern. From the beginning, the Centers
for
Disease Control and Prevention (CDC) has been continuously updating
the progress of COVID-19 and the status of public health safety in
the US in their Morbidity and Mortality Weekly Report
(\url{https://www.cdc.gov/mmwr/index.html}).
A large body of literature has reported clinical characteristics of
the disease \citep[e.g.,][]{fauci2020covid,moghadas2020projecting,
	parag2020clinical}. For the general public, however, the case and
death counts at the nation and state levels are of direct concerns.
Although there were high volumes of confirmed and death cases
reported in the US in March and April, the spread of the disease
effectively slowed down in late April and early May owing to the
practice of social distancing, ``mask wearing'' recommendations,
``stay-at-home'' and anti-gathering orders, and many other quarantine
measures \citep[e.g.,][]{tian2020evaluate}. However, the number of
new confirmed cases in some of the US states has surged since early
June, two weeks after the resumption of business
in a majority of the states. In terms of the dynamic pattern of the
case counts, some states are more similar than others
possibly due to difference in quarantine and reopening policies.

We clustered the US states by the patterns in their COVID-19
case count series, which was treated as a times series clustering
problem. Such clustering analysis has not yet been reported in the 
literature
of the study for COVID-19.
\emph{Feature-based} algorithms are widely used for time series 
clustering,
where features are first generated from the raw data
and then appropriate clustering algorithms are applied to form
clusters \citep{liao2005clustering}.
Many studies on the combination of feature generating procedure
and clustering algorithms have been done.
For example, an anytime algorithm based on the Haar wavelet
decomposition followed by a modified $k$-means clustering was
proposed by~\citet{lin2004iterative}; an
agglomeration algorithm incorporated with principals
component analysis was introduced by \citet{shaw1992using}. An
appealing alternative to solving time series clustering problems is
to exploit functional data analysis (FDA) techniques, such as the
$k$-centers functional clustering
algorithm \citep{chiou2007functional} and an EM-based algorithm for
Gaussian mixture models~\citep{chen2015EMCluster}, the latter of
which has been in a COVID-19 application~\citep{tang2020functional}.
See~\citet{jacques2014functional} for a concise survey for
functional data clustering methods.

Our time series clustering was based on a \emph{non-negative matrix
	factorization} (NMF) followed by a $k$-means clustering 
	procedure.
The series of each state was properly approximated
by a linear combination of a small number of bases. The coefficients
of the bases were used as features in a $k$-means
clustering analysis. This NMF-based method is attractive for
its robustness, stability, and
scalability~\citep{brunet2004metagenes,devarajan2008nonnegative}.
In particular, we considered an extended
NMF method allowing for missing values in the dataset. The number of
bases was determined through a cross-validation method where,
facilitated by a stratification on the level of the data, a randomly
selected fold of data was treated as missing and predicted by the
rest. Our clustering results
provide a platform to further study the similarities of the states
that are in the same cluster and the dissimilarities of those that
are not. Further, we investigated the changes in the clustering
results as additional data became available with an
increment of one week successively. The temporal changes between the
consecutive analyses were explored via similarity measures to
identify possible change point(s) in the clustering of the states.
Our analysis result suggests that the policies varying among the
states may have impacted the spreading or confinement of the 
pandemic.

The rest of the manuscript is organized as follows. In
Section~\ref{sec:data}, we introduce the data source, basic
descriptive statistics, and data preprocessing procedures. In
Section~\ref{sec:method}, we elaborate the fundamental concept of
NMF, and propose an NMF-based algorithm followed by $k$-means
clustering. The proposed algorithm is applied to the preprocessed
data and the clustering results for different study periods are
presented in Section~\ref{sec:result}. The temporal dynamic of
cluster structures is investigated in Section~\ref{sec:result} as
well. Lastly, we address some concluding remarks and carry out some
discussions in Section~\ref{sec:disc}.

\section{Data}
\label{sec:data}

Daily new confirmed cases from each US state were retrieved from a
public repository maintained by the Center for Systems Science and
Engineering at the Johns Hopkins University
\citep{dong2020interactive}. The start date of our study period was
set to be Sunday (March 22, 2020). President
Trump declared a national emergency concerning the COVID-19 outbreak
on March 13, 2020, after which the number of tests increased
substantially. On average, it took about a week to get the results of
nasal and swab tests. Hence, few confirmed cases were reported in,
for example, the District of Columbia and West
Virginia, before March~22. The end date of our study period was
Saturday, July~25, 2020, when the spread of the pandemic appeared to
be slowing down in those states affected largely by the second wave.
The entire study period has $126$ days.

We considered 49 continental state-level entities (including the
District of Columbia), hereafter referred to as 49 states for
simplicity. Alaska and Hawaii were not included because the
population movements between the two states and the continent were
rather limited. Since the 49 states vary a lot in population size,
the raw counts of cases are not comparable across them.
Thus, we standardized the raw counts by the population estimates of
the states at the end of 2019 from The US Census Bureau
(\url{https://www.census.gov}). Specifically, the
standardized number of daily new confirmed cases for each state
was recorded in terms of ``per million'' over the study period.

\begin{figure}[tbp]
	\centering
	\includegraphics[width=\textwidth]{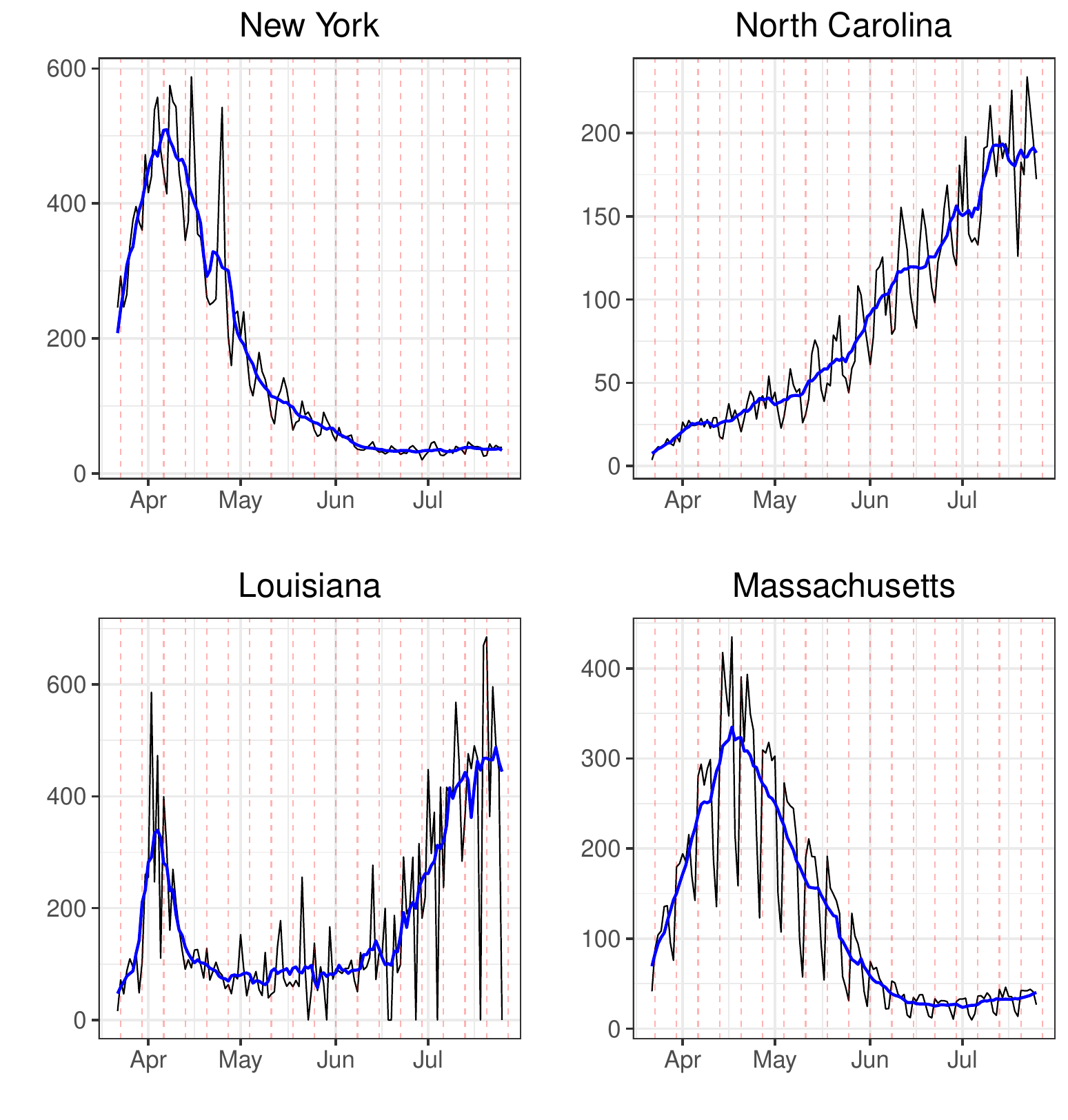}
	\caption{Daily confirmed cases in New York, North Carolina,
		Louisiana and Massachusetts over the study period, where the
		periodicity is reflected in vertical red dashed lines (7 
		days per
		gap).}
	\label{fig:periodic2}
\end{figure}

Figure \ref{fig:periodic2} displays the raw daily new confirmed cases
in four states with drastically different patterns ---
New York, North Carolina,
Louisianan and Massachusetts --- over the study period.
Periodicity was observed in the series with a period of 7 days.
The vertical red dashed lines in Figure \ref{fig:periodic2} are
presented with 7-day gaps.
The 7-day period was possibly caused by the routine of tests 
available
on different business days and testing report
delivery by the public health departments. In each panel, the thick
blue curve shows the 7-day {\em moving average} (with boundary
adjustment) of the new
confirmed cases per million in the state. The moving averages are
much smoother than the raw counts, allowing us to observe the
long-term trend in the time series and to make comparisons among the
states more easily. The moving average series are the input data for
our clustering analysis.

\section{Methods}
\label{sec:method}

Let $(x_{ij})$ be the input data matrix, $i = 1, 2, \ldots, 49$, $j
= 1, 2, \ldots, 126$, where $x_{ij}$ denotes the number
of confirmed cases per million (7-day moving average) of state~$i$ on
day~$j$. With the starting date held on March 22, 2020, clustering 
can
be done on each sub-period with incrementally extended ending date, 
in
addition to the whole study period. The collected analyses on these
sub-periods may help identifying an increment in time period that
might have led to
substantial changes in the clustering results. For a given time
period, our clustering is carried out on the coefficients of NMF
bases of the series. A potential change point in the cluster 
structure
is suggested when the consistency measures of every two consecutive
clustering outcomes are too low.
The components of our method are presented next.

\subsection{NMF}
\label{sec:nmf}

NMF is a class of matrix factorization methods where
a high-dimensional non-negative matrix is approximated
by the product of non-negative low-rank matrix factors. Consider
an $n\times m$ non-negative
matrix $\bm{X} := (x_{ij})$, where the rows and columns respectively
represent observations and features. In our data analysis, we have
$n = 49$ rows, each representing the time series of a state, and
$m$ columns, which is the number of days in the study period, each
representing one day of all the states in the time series.
An NMF of $\bm{X}$ determines
non-negative matrices $\bm{W} := (w_{ij})$ and $\bm{H} := (h_{ij})$
such that
\begin{equation}
	\label{eq:nmf}
	\bm{X} \approx \bm{W} \bm{H},
\end{equation}
where $\bm{W}$ and $\bm{H}$ are, respectively, $(n \times r)$ and
$(r \times m)$ matrices with $r < m$ being the number of NMF bases
of $\bm X$.
In essence, each row of~$\bm{H}$ is a basis vector of dimension $m$;
each row of $\bm{X}$ is approximated by a linear combination of these
bases, where the coefficients are stored in the rows of $\bm{W}$.
A good approximation of $\bm{X}$ indicates that most of the
variation in~$\bm{X}$ is captured by the resultant linear
combination of the bases.
The number of bases $r$ is usually much smaller than $m$.
This dimension reduction procedure allows us
to explore the $(n \times m)$ matrix $\bm{X}$ more
efficiently through a low dimensional matrix
$\bm{W}$.
That is, the series of state~$i$ is represented by a coefficient
vector of dimension $r$ of the bases in the $i$-th row of $\bm{W}$.

The algorithm that we used specifically was the
\emph{weighted non-negative factorization} (WNMF) algorithm
\citep{guillamet2001aweighted}, which, as to be shown in the next
subsection, facilitates the selection of the tuning parameter~$r$.
Let $\bm{V} := (v_{ij})$ be a weight matrix of the same
dimension as $\bm{X}$, where $v_{ij}$
is the weight of $x_{ij}$, reflecting its relative importance.
The matrix approximation problem accordingly becomes
\begin{equation}
	\label{eq:wnmf}
	\bm{V}\odot\bm{X} \approx \bm{V}\odot(\bm{W} \bm{H})
\end{equation}
subject to the constraints of $w_{ij}, h_{ij} \geq 0$ for all $i, j$,
where $\odot$ denotes the operator of element-wise product. When
all the elements in $\bm{V}$ are ones, WNMF is reduced to the
standard NMF.

Consider a squared error cost function
\begin{equation}
	\label{eq:squareddiswnmf}
	\|\bm{V} \odot (\bm{X} - \bm{WH}) \|^2= \sum_{i = 1}^{n} \sum_{j 
	=
		1}^{m} v_{ij} \left(
	x_{ij} - \bm{w}^{\top}_i \bm{h}_j \right)^2,
\end{equation}
where $\bm{w}^{\top}_i$ and $\bm{h}_j$ are respectively the $i$-th
row of $\bm{W}$ and the $j$-th column of $\bm{H}$.
\citet{lee2000algorithms} developed a multiplicative update rule that
would minimize the cost function for NMF by iteratively updating
$\bm{W}$ and $\bm{H}$. \citet{wang2006ls} extended the
multiplicative rule such that it would be applicable to WNMF. Let $t
\in \mathbb{N}$ index the
iteration process. At the $t$-th iteration, the updates
of $\bm{W}^{(t + 1)}$ and $\bm{H}^{(t + 1)}$ from $\bm{W}^{(t)}$ and
$\bm{H}^{(t)}$ are given by
\begin{align}
	\label{eq:updateVH}
	\bm{H}^{(t+1)} \leftarrow
	\bm{H}^{(t)}\odot\frac{{\bm{W}^{(t)}}^\top(\bm{X}\odot\bm{V})}
	{{\bm{W}^{(t)}}^\top[(\bm{W}^{(t)}\bm{H}^{(t)})\odot\bm{V}]},
	\\
	\label{eq:updateVW}
	\bm{W}^{(t+1)} \leftarrow
	\bm{W}^{(t)}\odot\frac{(\bm{X}\odot\bm{V}){\bm{H}^{(t)}}^\top}
	{[(\bm{W}^{(t)}\bm{H}^{(t)})\odot\bm{V}]{\bm{H}^{(t)}}^\top},
\end{align}
where $\bm{X}^{\top}$ represents the transpose of $\bm{X}$. The
alternating update goes on
until the absolute difference between the maximum and the minimum
costs over a number of successive iterations is below a preset
tolerance. Equations~\eqref{eq:updateVH} and~\eqref{eq:updateVW}
jointly guarantee that Equation~\eqref{eq:wnmf} is non-increasing,
and, consequently, that the algorithm at least
converges to a local minimum \citep{lee2000algorithms}. We used
the WNMF algorithm implemented in \proglang{R} package \pkg{NMF}
\citep{gaujoux2010aflexible}.

Like any iterative algorithm, WNMF needs starting matrices
$\bm{W}^{(0)}$ and $\bm{H}^{(0)}$. We used the nonnegative double
singular value decomposition \citep{boutsidis2008svd} to initialize
the WNMF algorithm. In its basic form, this method selects the
starting matrices that approximate the singular value decomposition
of $\bm{X}$ by dropping the non-positive singular values and
replacing the negative entries in the unit-rank matrices formed by
singular vectors with zeros. The algorithm does not grant
randomization, which ensures reproducibility. It rapidly provides
starting matrices with error almost as small as those from competing
initialization approaches.

\subsection{Rank Selection for NMF}
\label{sec:basis}

It is critical to select an appropriate rank or the number of bases
for NMF. We selected~$r$ through a cross-validation-based
scheme extended from the method proposed by \citet{kanagal2010rank}.
For an $s$-fold cross-validation, the entries in $\bm{X}$ are
partitioned randomly into $s$ sub-groups or folds. For a given~$r$,
each of fold is held out as testing data, while the remaining
$(s - 1)$ folds are used as training data with the held-out entries
regarded as missing values to run a NMF. The mean squared
prediction error (MSPE) of each held-out fold is computed by 
comparing
the observed value with predictions using the NMF based on the
training data. The total MSPE summed over all folds is used as the
cross-validation criterion.

NMF for matrices with missing entries can be approached by
a WNMF problem \citep{kim2009weighted}. Let
the weight matrix $\bm{V}$ indicate the missing values in the
held-out fold:
\begin{equation}
	\label{eq:vij}
	v_{ij} =
	\begin{cases}
		1, \qquad & \mbox{if } x_{ij} \mbox{ is observed};\\
		0, \qquad & \mbox{if } x_{ij} \mbox{ is missing}.
	\end{cases}
\end{equation}
The missing values can be predicted by the corresponding values in
the approximation matrix $\bm{W}\bm{H}$ after implementing WNMF.
Given a candidate set of $r \in \{2, 3, \ldots, 12\}$, where the
maximum candidate value was about a quarter of the number of the
states, the one with the minimum MSPE was selected for the value
of~$r$.

The right-skewness of the COVID-19 case counts in most states needs
some care in WNMF. The data magnitudes in the right columns (later
time) are much larger than those in the left (earlier time). A random
partitioning of ${\bm X}$ may cause the data distribution in the
testing fold significantly different from the counterparts in the
training folds, potentially resulting in poor predictions. Having
this in mind, we considered a stratified partitioning scheme.
Specifically, all the entries of $\bm{X}$ were stratified into two
strata based on the magnitudes, and a simple random sampling was
executed within each stratum. The split of the two strata was set at
the $75$-th percentile of the observed data in the study period,
giving the most stable results based on our experiments.

\subsection{Clustering Procedure}
\label{sec:step}

Our clustering procedure based on WNMF has two steps, the first of
which is dimension reduction with NMF. This step starts with rank
selection of NMF as presented in the last subsection. For the 
selected
rank $r$,
\begin{equation}
	\label{eq:nmfrow}
	\bm{x}_i^{\top} \approx \bm{w}_i^{\top} \bm{H},
\end{equation}
where $\bm{x}^{\top}_i$ is defined analogously as $\bm{w}^{\top}_i$.
Equation~\eqref{eq:nmfrow} implies that the ($m$-dimensional)
features of the $i$-th subject in $\bm{X}$ are well approximated by
the ($r$-dimensional) corresponding coefficients in $\bm{W}$,
rendering a cluster analysis through $\bm{W}$.

The second step is to perform a clustering analysis using the basis
coefficients $\bm{W}$ as input. A variety of well-developed 
clustering
algorithms can be adopted, and our choice was the $k$-means
clustering as available in \proglang{R} 
\citep{hartigan1979algorithm}.
Different combinations of basis coefficients represent different
groups in both the pattern of the times series and their magnitudes.
The algorithm depends on a random initialization.  We used 500 
random starting point and chose the result that gave the smallest
within-cluster sum of squares (WSS).
To select the number of clusters~$g$, we used the elbow
method which plots the WSS as a function of~$g$, and
chose~$g$ as the elbow of the curve.
With 49 states, we considered candidate $g \in \{2, 3, \ldots, 10\}$.
The determination of the elbow position may vary from case to case.
Our selection considered both the curvature and the interpretation
of the clustering results.

\begin{algorithm}[tbp]
	\label{alg:wnmf}
	\caption{Pseudo-algorithm of the NMF-based $k$-means
		clustering procedure for the daily COVID-19 case counts in
		49 states.}
	\KwIn{Raw COVID-19 daily COVID-19 case counts in 49 states}
	\KwOut{Clustering results of the 49 states}
	Scale by population and apply $7$-day moving average for each
	state to obtain $\bm{X}$\;
	{\For{$r = 2 $ {\rm to} $12$} {
			Carry out cross-validation based on WNMF of $\bm{X}$ 
			with rank $r$\;
			Compute ${\rm MSPE}_r$\;}
	}
	Select NMF rank {$r^{\rm (opt)} = \argmin_r \{{\rm MSPE}_r\}$}\;
	Obtain $\bm{W}$ based on NMF rank $r^{\rm (opt)}$\;
	\For{$g = 2 $ {\rm to} $10$} {
		Carry out $k$-means for $\bm{W}$ with $g$ clusters\;
		Compute WSS for this $g$, ${\rm WSS}_g$\;
	}
	Select the number of clusters
	$g^{\rm (opt)}$ based on the elbow plot and practical
	interpretation\;
	\Return{} Result of $k$-means clustering for $\bm{W}$ with
	$g^{\rm (opt)}$ clusters\;
\end{algorithm}

Algorithm~\ref{alg:wnmf} summarizes the major steps of the
NMF-based $k$-means clustering algorithm for the daily COVID-19 new
case counts from 49 states in a given time period.

\subsection{Temporal Dynamics in Cluster Structure}
\label{sec:change}

Algorithm~\ref{alg:wnmf} can be applied to the new case count series
of any time period, which facilitates an investigation of how the
clustering results change over time as more data becomes available.
We fixed the starting date at March~22, 2020 and considered a 
sequence
of time periods with ending date starting from March~28, 2020, with 
an
increment of one week until July~25, 2020. The incremental
segment of 7 days is chosen according to the observed 7-day period
in the data; see Figure~\ref{fig:periodic2}. Such arrangement led to
a total of 18 study periods. The clustering results over the 18
periods with incremental ending dates provided information about the
temporal dynamics of the cluster structure.

At each ending date, we compared the obtained  clustering results
with or without the data from the most recent week. The
agreement between the two clustering results was assessed by
\emph{adjusted rand index} \citep[ARI,][]{hubert1985comparing}.
The ARI measurement was originally proposed to assess the accuracy 
of a
clustering strategy by comparing its result with the ground truth. 
ARI
ranges between $-1$ and $1$,
with higher values indicating that the clustering result is closer to
the truth. We use it here to quantify the degree of agreement
between a pair cluster results. A small value of ARI suggests
that the two clustering results are quite different from each other.
A big drop after a sequence of high values may indicate a
structural change in the clustering results.

\section{Results}
\label{sec:result}

We first report the clustering results on the
entire study period (from March~22 to July~25, 2020), then
investigate the temporal dynamics in the clustering
results, and finally compare the cluster structures before and after
a possible change point.

\subsection{Clustering Result for the Entire Study Period}
\label{sec:resultentire}

\begin{figure}[tbp]
	\includegraphics[width=\textwidth]{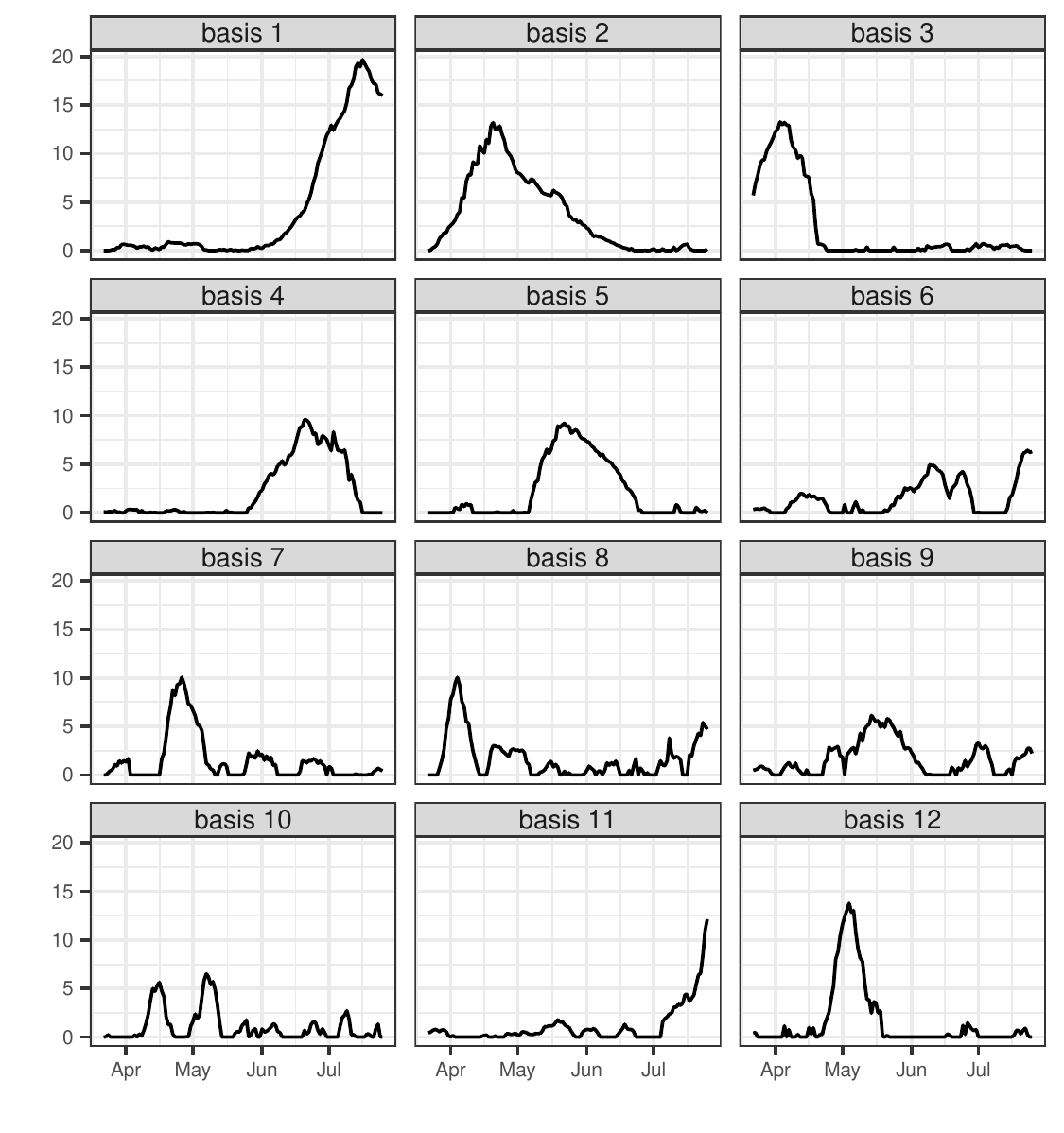}
	\caption{NMF bases for the proposed algorithm applied to the time
		series data from March 22 to July 25, 2020.}
	\label{fig:basis2}
\end{figure}

Algorithm~\ref{alg:wnmf} was applied to the preprocessed time series
data for the entire study period from March~22 to July~25, 2020. The
NMF step yielded 12~bases as displayed in Figure~\ref{fig:basis2}.
Most of the bases show distinctive
features of peaks in terms of their timings and shapes.
In chronological order, we see clearly a series of bases peaking
with a size of about~10 or higher from early April to late July.
The bases have different patterns in addition to their sizes.
For example, basis~2 and~3 with early peaks are important for
capturing high counts in early stage of the study period. Basis~2 is
about a week later but goes down much more slowly than basis~3;
basis~3 spans about 2~weeks while basis~2 spreads for about 5~weeks.
Basis~1 and basis~11 with later peaks are important for capturing 
high
counts in later stage of the study period, with basis~1 having an
earlier climbing point and earlier peak than basis~11.
Bases with less obvious interpretations and lower magnitude help
properly adjust the
ups and downs in describing the daily count curves of the 49 states.

\begin{table}[tbp]
	\caption{Summary of the clustering results based on
		Algorithm~\ref{alg:wnmf} for the entire study period (March 
		22
		to July 25, 2020)}
	\centering
	\setlength{\tabcolsep}{3pt}
	\begin{tabular}{c  m{0.9\textwidth}}
		\toprule
		Cluster & {\centering States} \\
		\midrule
		\rowcolor{beaublue} \textbf{A}  &   { New Jersey,  New
			York}  \\
		\rowcolor{bananamania} \textbf{B}       &
		{ Connecticut, Delaware, Massachusetts,
			Rhode Island } \\
		\rowcolor{beaublue} \textbf{C}       &
		{ District of Columbia,
			Illinois, Iowa,
			Maryland, Minnesota, Nebraska, Virginia} \\
		\rowcolor{bananamania} \textbf{D}       & {
			Arizona} \\
		\rowcolor{beaublue} \textbf{E}       & {
			Louisiana}
		\\
		\rowcolor{bananamania} \textbf{F}       & {
			Alabama, Arkansas,
			California, Florida,
			Georgia, Idaho,
			Mississippi, Nevada,
			North Carolina,
			South Carolina,
			Tennessee, Texas,
			Utah
		} \\
		\rowcolor{beaublue} \textbf{G}       & {
			Colorado,
			Indiana,
			Kansas, Kentucky,
			Maine, Michigan, Missouri,
			Montana, New Hampshire,
			New Mexico, North Dakota,
			Ohio, Oklahoma, Oregon,
			Pennsylvania, South Dakota,
			Vermont,
			Washington, West Virginia, Wisconsin, Wyoming} \\
		\bottomrule
	\end{tabular}
	\label{tab:state}
\end{table}

\begin{figure}[tbp]
	\centering
	\includegraphics[width=\textwidth]{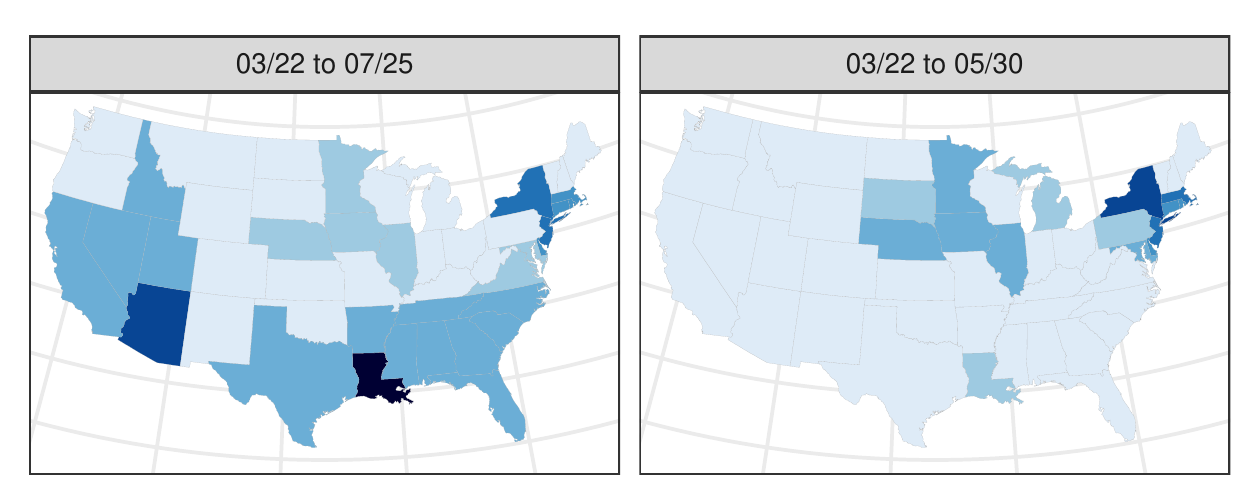}
	\caption{Clustering result for the entire study period (from 
	March
		22 to July 25, 2020) in the left panel; clustering result 
		for the
		first
		half of the entire study period (from March 22 to May 30, 
		2020,
		where May 30, 2020 appears to be a potential change point of
		dynamics) in the right panel.}
	\label{fig:map2}
\end{figure}

\begin{figure}[tbp]
	\includegraphics[width=\textwidth]{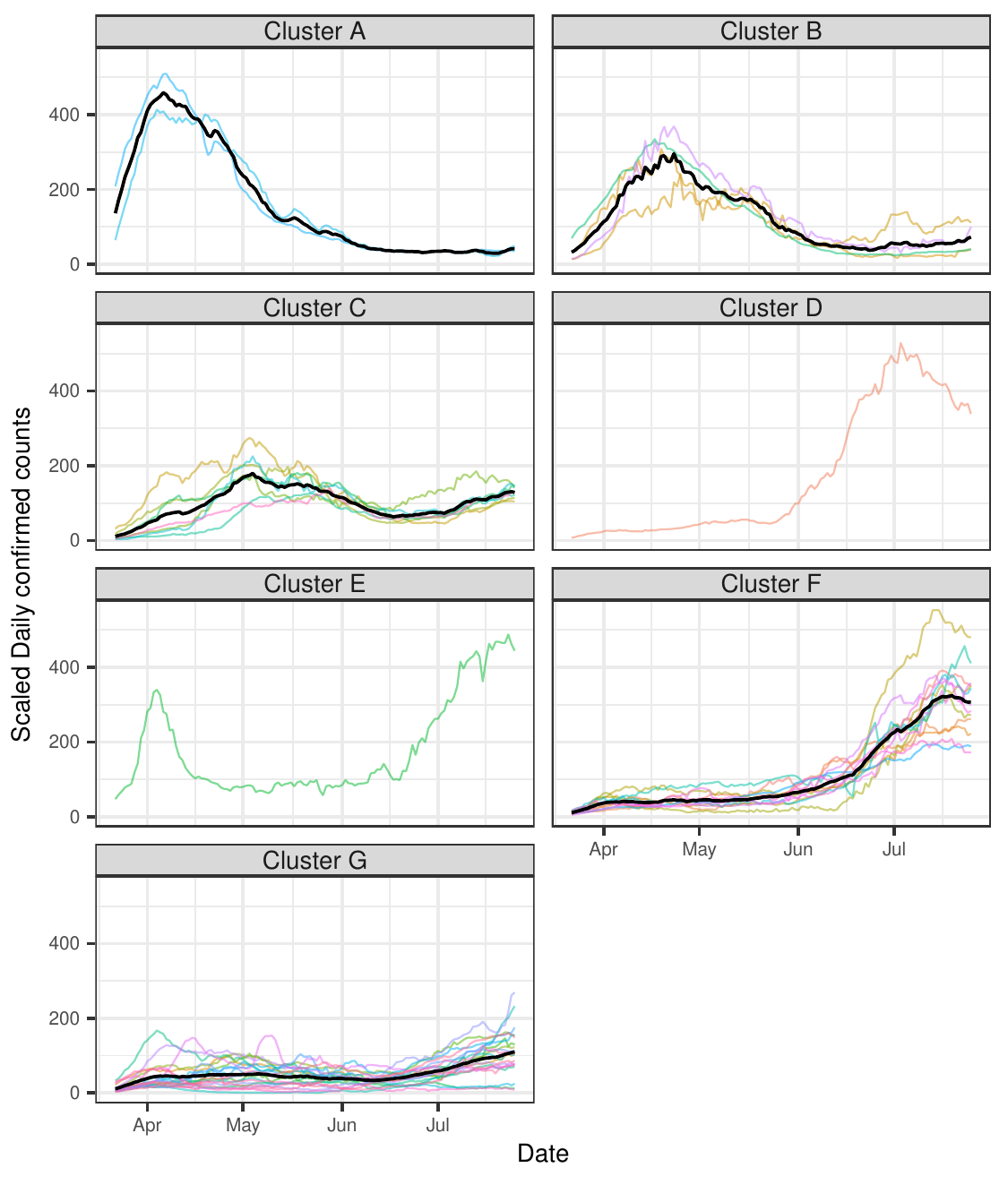}
	\caption{Daily confirmed case counts (after scaling and 
	smoothing)
		of the states in each cluster; the mean curves are presented 
		by
		black thick curves. Especially, each of the clusters D and E 
		has
		one state member only.}
	\label{fig:curves}
\end{figure}

The $k$-means step on the coefficients of the 12~bases grouped the
49~states into 7~clusters as summarized in Table~\ref{tab:state},
with a graphic visualization presented in the left panel of
Figure~\ref{fig:map2}. The means of the daily new case series per
million (in population) for the 7~clusters are overlaid with the
series themselves in each cluster in
Figure~\ref{fig:curves}. As expected, the curves in the same cluster
appear similar in shape, but present different patterns or trends
across the clusters. The states that are geographically
close are likely to be in the same cluster due to the spillover
effect. Other factors such as mobility, transport capacity,
population density, and state-level social distancing measures may
have been in effect, too, as discussed next.

In the sequel, we use the two-letter state abbreviations to represent
the states for brevity in our discussion
(\url{https://www.ssa.gov/international/coc-docs/states.html}).

There are two singleton clusters D and E, which contains AZ
and LA, respectively, presenting completely different
characteristics of the new case count
curves as shown in Figure~\ref{fig:curves}. The curve of AZ was flat
early
on, but started climbing at the beginning of
June with a peak in the first week of July. Although a declining
trend was observed since then, the counts had remained high until
the end of the study period. The coefficient of basis~4 for AZ is
outstanding, while the coefficient of basis~1 is relatively large as
well, together precisely presenting the curve feature. The observed
curve pattern may have to
do with the state policies. The ``stay at home'' order in AZ was
in effect from March~31 to May~15, and a second ``business
closing'' order was
issued on June~29 in response to the fast growing confirmed cases in
June. On the other hand, the LA curve has two peaks, one in early
April and the other in late July. It is the only bimodal curve in
Figure~\ref{fig:curves}. The coefficients of basis~1 and 3 for LA
are both relatively greater than the counterparts of most of the
other states, capturing the observed bimodal pattern. In fact, LA
was one of the states with the highest number of confirmed cases
per capita among all of the US states in July and August
({\url{https://covid.cdc.gov/covid-data-tracker/#cases_casesper100k}}).
The first outbreak lasted from late March to mid April,
possibly
caused by the Mardi Gras festivities in late February
when millions of people traveled to New Orleans. The second wave
started in late June, following the phase~2 of business reopening.
Both peaks in LA are higher than the peaks in most of the other
states.

The two states in Cluster~A are NY and NJ whose new count case
curves resemble each other throughout the study period. The COVID-19
broke out in NY and was rapidly transmitted to its neighbor NJ in
mid March as a large number of people working in New York City (NYC)
were resident in northern NJ across the Hudson river.
The early counts of daily new cases in NY and NJ were
significantly higher than all of the other continental states. After
peaking in early April, the pandemic appeared to be
well-controlled owing to the practice of strict (mandatory) public
health policies; for instance, the business shutdown in NYC started
in mid March and the face covering order was issued on
April~15. The number of new cases went down and did not
bounce back until July. Hence, the NY and NY curves are completely 
opposite to
the AZ curve, reflected in relatively large coefficients of bases~2
and~3.

The four states in Cluster~B, CT, DE, MA and RI, are all
geographically close to NY and NJ. Their mean curve is similar to
that of Cluster~A in shape; both are unimodal and the peaks appear
in the early times. The main difference among the curves in
Clusters~A and~B is reflected in magnitude; the NY and NJ curves (in
Cluster~A) have been consistently higher over time. A minor
difference is that the peaks in NY and NJ both occurred in early
April, compared to the peaks of the four states occurring around
late April. The four state curves did not show any tendency of
rebound since May until the end of the study period, either. The
similarity among the curves in Clusters~A and~B may be attributed to
the similar state government policies. All of the six states
are participating parties of the Eastern States Multi-State
Council which, as announced on April~13, 2020, worked together to
develop a fully integrated regional framework to gradually lift the
states' ``stay at home'' orders while minimizing the risk of
increased spread of the virus
(\url{https://en.wikipedia.org/wiki/Eastern_States_Multi-state_Council}).
The low counts since June appeared to support the effectiveness of
their collective efforts.

Cluster~C contains 7~states, DC, IA, IL, MD, MN, NE and VA.
Their mean curve is flatter than those of Clusters~A and~B,
and the values are smaller in magnitude in the early and mid times.
The peak of the average occurred in early May, followed by a decline
until a second
increasing period that started in July.
Three of them (DC, MD and VA) belong to the Washington
metropolitan area. Similar government orders were
issued at about the same time in this region. For example, public
schools were closed in mid March in DC and VA; the ``stay at home''
orders were issued on March~30 by all three state governments.
The other four states are contiguous midwest states.
Chicago in IL and Omaha in NE were hot
spots during the outbreak early on. In particular, Chicago
is one of the biggest ports of entry in the US; both Chicago and 
Omaha
have fully functional interstate transportation facilities. The high
mobility may have contributed to their high case counts in May.

Cluster~F contains 13 states with low counts of new cases from the
start of the study period to June, followed by a continuing rise
in June and July. The states form two big blocks respectively in the
west (CA, NE, ID and UT) and south (AL, AR, FL, GA, MS, NC, SC, TN
and TX).
Among them, CA was one of the first states reporting confirmed
cases, with several
counties marked as hot spots. Although CA is a populated state,
the pandemic was effectively controlled due to the rapid
and powerful response by the state government. In March, CA became
the first state to announce business shutdown, which was kept in
effect until early May. Several populated counties such as San
Francisco, Santa Clara, and Contra Costa even ordered
residents to shelter in place for three weeks for the purpose of
slowing down the
mobility. Most of the other states in this cluster had similar orders
in late March. The timely public health measures worked well to keep
the curves flat at the early and mid times. On the other hand,
the increased population mobility in May might
have contributed to the rapidly increased case counts during the
later periods.
Most of the states announced business reopening in late April or 
early
May. In addition, six of the top ten busiest airports
are located in CA, GA, FL, NV and TX.  It is worthy of
mentioning that the AZ curve is similar to the overall trend of the
curves in Cluster F. However, the AZ curve is separated in a
singleton group, possibly because it started increasing at the
beginning of June, about two weeks prior to the climbing trend
observed in the states in Cluster~F.

The remaining 21 states are in Cluster~G. Overall, these states have
relatively low counts of new case every day, and their curves are
flat throughout the study period. States in this cluster have either
low population density or mandate strict quarantine measures.
For instance, WY, MT, ND and SD are four of the top five states with
the lowest population densities.
The first domestic confirmed case of the COVID-19 was reported in
Seattle, WA on January~22, and the King county (in Seattle) became
one of the earliest hot spots in the US. In response to the outbreak,
the state governor declared state emergency as early as February~29,
and shut down the schools statewide on March~3. A series of
instructions and orders related to mitigating the impact of the
COVID-19 were proposed and executed subsequently.
The business reopening
plan in WA was not implemented until May~26, later than most of the
other states (including the states that were much more severely hit
by the coronavirus) in the US. All those efforts may have caused that
the counts of daily confirmed cases in WA started decreasing since
the second week of April, and remained at a low level until the end
of the study period.

\subsection{Temporal Dynamic of Cluster Structure}
\label{sec:clusterchange}

In this section, we look into the temporal dynamics in cluster
structure. The proposed clustering method was applied to 18 sub
study periods (all starting from March~22) with end point
successively incremented by one week,
i.e., April~4, April~11, and so on. Let $T_1, T_2, \ldots, T_{18}$
indicate the 18 cumulatively incremental study periods in sequence.
Table~\ref{table:param} summarizes the rank of NMF (denoted $r$) and
the
number of clusters (denoted $g$) of each study period. The number of
bases~$r$ increases in general as the duration of the study extends.
This is expected as longer study period usually admits more complex
curve patterns which require more bases to approximate.
The number of clusters~$g$ started with~5 increased a little over
time, and remained at~7 since $T_{15}$.

\begin{table}[tbp]
	\caption{The selected tuning parameters for 18 study periods
		with end point incremented by one week.}
	\centering
	\begin{tabular*}{\textwidth}{l@{\extracolsep{\fill}}
			lccccccccc}
		\toprule
		Study period & $T_1$ & $T_2$ & $T_3$ & $T_4$ & $T_5$ & $T_6$
		& $T_7$ & $T_8$ & $T_9$ \\
		\arrayrulecolor{black!30}\midrule
		Rank of NMF ($r$)  & 2 & 5 & 4 & 6 & 6 & 8 & 7 & 8 & 8 \\
		Number of clusters ($g$)  & 5 & 5 & 5 & 6 & 5 & 5 & 6 & 6 & 
		6  \\
		\arrayrulecolor{black}\midrule
		Study period & $T_{10}$ & $T_{11}$ & $T_{12}$ & $T_{13}$
		& $T_{14}$ & $T_{15}$ & $T_{16}$ & $T_{17}$ & $T_{18}$ \\
		\arrayrulecolor{black!30}\midrule
		Rank of NMF ($r$)  & 7 & 8 & 8 & 12 & 11 & 12 & 11 & 12 & 12 
		\\
		Number of clusters ($g$)  & 6 & 7 & 7 & 8 & 6 & 7 & 7 & 7 & 
		7 \\
		\arrayrulecolor{black}\bottomrule
	\end{tabular*}
	\label{table:param}
\end{table}

\begin{figure}[tbp]
	\centering
	\includegraphics[width=\textwidth]{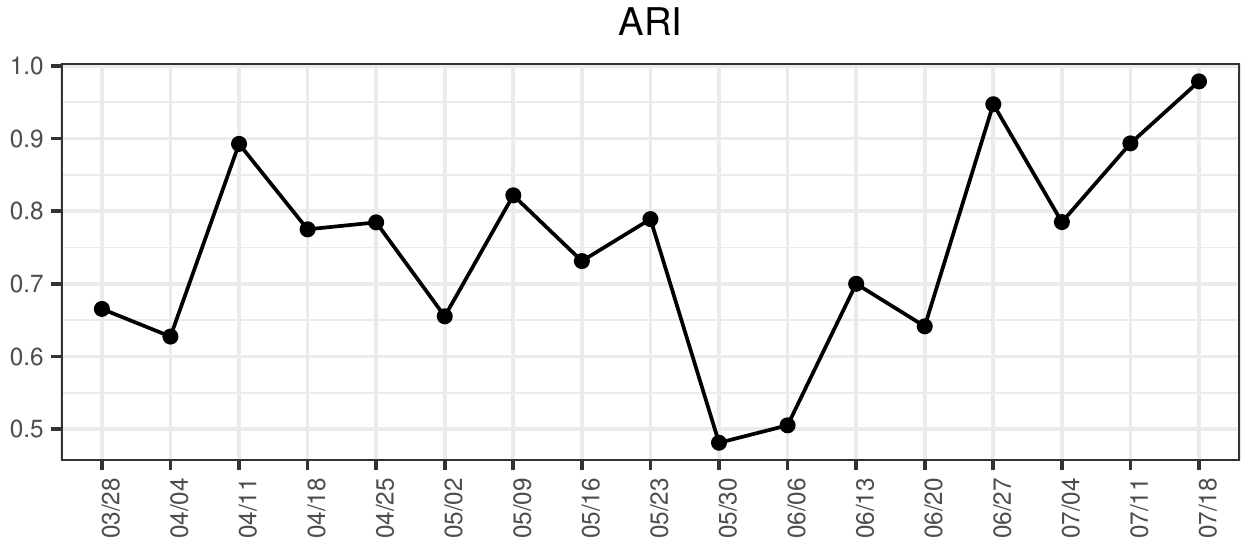}
	\caption{ARI for successive pairwise clustering results as
		the study period was extended by one week.}
	\label{fig:ARI}
\end{figure}

To identify possible structural changes in the clustering results,
we checked the consistency in the resultant clusters as measured by
ARI before and after each week (i.e., $T_1$ versus $T_2$,
$T_2$ versus $T_3$ and so on).
Figure~\ref{fig:ARI} presents the ARI for each
successive comparison between the clustering results from $T_i$
versus those from $T_{i + 1}$, $i = 1, 2, \ldots, 17$, where the
horizontal axis represents the end point of the shorter time
period in each pair.
ARI has a moderate value at the very
beginning, and vibrates in the early few
weeks. A tremendous drop in $T_{10}$ heralds a
large degree of inconsistency. Hence, possible structural change
in the clustering results may occur during the one-week periods
before and
after May~30. ARI remains relatively low until a
rising in $T_{12}$ as the week of
June~20 is included. After a slight drop at $T_{13}$, ARI stays at a 
high
level until the end of the study period.

The possible change in the clustering results as the study period up
to the week of May~30
is appended echoes the findings in the clustering patterns in
Figure~\ref{fig:curves}. It was during the period from May~30 to
June~20 that the rebound in many of the states in Clusters~C, E, and
G and the rise in the states in Clusters~D and F occurred. Noticing
the estimated mean of incubation period is around 5.34 days with
95\% confidence interval $[4.29, 6.40]$ \citep{zhang2020meta} and
that the majority (97.5\%) of the infected individuals developed
symptoms within 11.5 days \citep{lauer2020incubation}, we
speculated that
the actual change point was in the week between May~17 and May~23.
This week coincides with the timing of business reopening in most of
the states in the US; for instance, the second phase of reopening in
Texas was on May~18, 2020
(\url{https://www.dshs.state.tx.us/coronavirus/opentexas.aspx}) and
the phase one reopening of certain sectors of the economy in 
Connecticut
took place on May~20, 2020
(\url{https://portal.ct.gov/DECD/Content/Coronavirus-Business-Recovery/}).
It is likely that the large-scale business reopening across the 
nation
had an association with the change of spread pattern of COVID-19.

\subsection{Comparison with the Results from Data Before May 30}
\label{sec:comparison}

The study of the temporal dynamic of cluster structure in
Section~\ref{sec:clusterchange} induces a comparison between
the clustering results including and excluding the data after
May~30, respectively corresponding to $T_{18}$ and $T_{10}$
(March~22 to May~30, 2020). As
shown in Table~\ref{table:param}, there are $6$ clusters for
$T_{10}$, where the detailed clustering result is summarized in
Table~\ref{tab:state0530}, followed by a graphic visualization given
in the right panel of Figure~\ref{fig:map2}.

Specifically, the ARI score between the clustering results
of $T_{18}$ and $T_{10}$ is 0.36.
The primary difference between the two clustering results is
that Clusters~D (AZ), F, and G (the majority) from $T_{18}$ are
grouped in one big
cluster F* from $T_{10}$. As seen in
Figure~\ref{fig:curves}, the curves in Clusters~D, F, and G are
all flat over $T_{10}$ in spite of several minor humps up to May~30.
Secondly, LA, MI, PA, and SD are in one cluster (Cluster E*) from
$T_{10}$, but from $T_{18}$,
LA forms a singleton and MI, PA and SD belong to Cluster~G.
The four states had early
peaks between April~7 and April~20, but the counts were
significantly lower than that from NY, rendering them distinct from
all the other states from $T_{10}$.
Thirdly, NJ and NY are in the same cluster from $T_{18}$,  but from
$T_{10}$, NY itself forms a single-state cluster, and  NJ and MA
together form another cluster. Different from NY that increased from
the beginning to the first week of April and then started declining
right after, NJ had a long peak for almost two weeks in April. In
addition, the magnitudes of the curves of NJ and MA were relatively
close to each other throughout $T_{10}$, consistently smaller than
that of NY. Cluster~D* from $T_{10}$ and Cluster~C from $T_{18}$ are
similar, with the exception of VA in C but not in D*.

\begin{table}[tbp]
	\caption{Cluster details of the period from 03/22/2020 to
		05/30/2020}
	\centering
	\setlength{\tabcolsep}{3pt}
	\begin{tabular}{c  m{0.9\textwidth}}
		\toprule
		Cluster & {\centering States} \\
		\midrule
		\rowcolor{beaublue} \textbf{A*}       &{New York}  \\
		\rowcolor{bananamania} \textbf{B*}       &
		{New Jersey, Massachusetts }\\
		\rowcolor{beaublue} \textbf{C*}       &
		{ Connecticut, Delaware,
			Rhode Island }
		\\
		\rowcolor{bananamania} \textbf{D*}       & {
			District of Columbia, Illinois, Iowa, Maryland,
			Minnesota, Nebraska} \\
		\rowcolor{beaublue} \textbf{E*}       & {
			Louisiana, Michigan,
			Pennsylvania, South Dakota}
		\\
		\rowcolor{bananamania} \textbf{F*}       & {
			Alabama, Arizona, Arkansas,
			California, Colorado, Florida,
			Georgia, Idaho, Indiana, Kansas, Kentucky,
			Maine, Mississippi, Missouri,
			Montana, Nevada, New Hampshire,
			New Mexico, North Carolina, North Dakota,
			Ohio, Oklahoma, Oregon,
			South Carolina,
			Tennessee, Texas,
			Utah, Vermont, Virginia,
			Washington, West Virginia, Wisconsin, Wyoming
		} \\
		\bottomrule
	\end{tabular}
	\label{tab:state0530}
\end{table}

A few factors might have caused the split of the largest
group G* from $T_{10}$ into Clusters~D,~F and~G from $T_{18}$. We
have closely looked into AZ in Cluster D in
Section~\ref{sec:resultentire}. The counts
in the states in Cluster~F are obviously
higher than those in Cluster~G on average after May~30.
The first factor may be the various state-level quarantine measures.
Almost all the state governments (in Cluster G) mandated business
shutdown (or
remote working) in March; however several highly populated states in
cluster~F, like FL, reopened the local business earlier than the
rest. Two other states (in Cluster F), TX and GA, also reopened
their business in early May, including
some entertainment facilities with massive gathering. Theaters and
clubs were reopen in GA in late April, while recreational trails and
beaches were reopen in FL on May~4. No face covering
orders were enforced for either indoor or outdoor activities in
those states. The second possible factor is population density.
Though some of the states in cluster~G announced reopening around
early May (e.g., CO, IN, ME, WY and WV), none of these states have
high population density regions like the Atlanta metropolitan area
in GA, the Miami-Fort Lauderdale-West Palm Beach metropolitan area
in FL or the Dallas-Fort Worth metroplex in TX.
Lastly, some states in cluster~F maintain a much busier traffic
systems which possibly cause larger volume of mobility.

\section{Discussion}
\label{sec:disc}

Our clustering procedure is a combination of NMF and $k$-means
clustering. NMF is used for dimension reduction and $k$-means
is applied to the basis coefficients from the NMF. NMF itself is a
clustering method in which the number of clusters is the rank and 
each
data point is assigned to a cluster corresponding to the most highly
expressed basis \citep{brunet2004metagenes}. When the basis are
restricted to be orthogonal, NMF and $k$-means are equivalent in 
terms
of the objective function \citep{ding2005equivalence}. Our method is
different from the traditional NMF clustering in that the clusters
are formed by the similarities in basis coefficients. It has the
advantage to capture patterns that are combinations of multiple 
bases,
which may not be appropriate to be clustered to the groups 
represented
by either one of the bases.
Our method is also preferred to the standard $k$-means clustering
because the $k$-means algorithm gives equal weights to all time
points. However, data of early time points may have a much smaller
magnitude. The
difference in early time times may not contribute as much as the 
later
times. As a result, states with distinctive
patterns at early times such as NY, NJ and LA, cannot be successfully
distinguished from other states by the standard $k$-means.

Our clustering of the 49 states appears to be reasonably reflecting
the spread and control of the pandemic in the US up to July 25, 2020.
The third wave since September as well as the differences in
quarantine and business resumption measures are expected reshuffle 
the
clustering reported in this paper, which merits continuing research.
The implementation of our method is publicly available at GitLab 
(\url{https://gitlab.com/covid-19-analysis/covid19-nmf.git}).

\end{document}